\documentclass[a4paper,manuscript]{aastex}

\usepackage{gensymb}

\defcitealias{parsons16}{P16}
\bibpunct{(}{)}{;}{a}{,}{,}{}

\shorttitle{Long-term spot stability in QS Vir}
\shortauthors{O. Latkovi{\' c} et al.}

\begin{document}

\title{Long-term spot stability in the post-common-envelope binary \\ QS Vir}

\author{Olivera Latkovi{\' c}}
\affil{Astronomical Observatory, Volgina 7, 11060 Belgrade 38, Serbia}
\email{olivia@aob.rs}

\author{Attila Cs{\' e}ki}
\affil{Astronomical Observatory, Volgina 7, 11060 Belgrade 38, Serbia}

\author{Gojko Djura{\v s}evi{\'c }\altaffilmark{1}}
\affil{Astronomical Observatory, Volgina 7, 11060 Belgrade 38, Serbia}
\altaffiltext{1}{Isaac Newton Institute of Chile, Yugoslavia Branch, Serbia}

\author{Ahmed Essam}
\affil{Department of Astronomy, National Research Institute of Astronomy and Geophysics, Helwan, Egypt}

\author{Amal S. Hamed}
\affil{Department of Astronomy, National Research Institute of Astronomy and Geophysics, Helwan, Egypt}

\and
\author{Shahenaz M. Youssef\altaffilmark{2}}
\affil{Department of Astronomy, National Research Institute of Astronomy and Geophysics, Helwan, Egypt}
\altaffiltext{2}{Department of Astronomy, Space Science and Meteorology, Faculty of Science, Cairo University, 12613 Cairo, Egypt}

\begin{abstract}
We observed the post-common-envelope eclipsing binary with a white dwarf component, QS Vir, using the 1.88 m telescope of Kotammia Observatory in Egypt. The new observations were analyzed together with all multicolor light curves available online (sampling a period of 25 years), using a full-feature binary system modeling software based on Roche geometry. This is the first time complete photometric modeling was done with most of these data. QS Vir is a detached system, with the red dwarf component underfilling its Roche lobe by a small margin. All light curves feature out-of-eclipse variability that is associated with ellipsoidal variation, mutual irradiation and irregularities in surface brightness of the tidally distorted and magnetically active red dwarf. We tested models with one, two and three dark spots and found that one spot is sufficient to account for the light curve asymmetry in all datasets, although this does not rule out the presence of multiple spots. We also found that a single spotted model cannot fit light curves observed simultaneously in different filters. Instead, each filter requires a different spot configuration. To thoroughly explore the parameter space of spot locations, we devised a grid-search procedure and used it to find consistent solutions. Based on this, we conclude that the dark spot responsible for light curve distortions has been stable for the past 15 years, after a major migration that happened between 1993 and 2002, possibly due to a flip-flop event.
\end{abstract}



\section{Introduction}

\object{QS Vir} is a binary system composed of a white dwarf and a low-mass M dwarf that nearly fills its Roche lobe, in a very tight orbit with a period of only 3.6 hours. Systems such as this are believed to be in the post-common-envelope phase of binary evolution and on the way to become semidetached cataclysmic variables, through further loss of angular momentum and shrinking of the orbit due to gravitational waves or magnetic breaking \citep[see][]{kraft62, paczynski67, paczynski76, verbunt81}. Thus, QS Vir can be classified (and is interchangeably referred to in literature) as either a post-common-envelope binary (PCEB) or a pre-cataclysmic variable (pre-CV).

PCEBs are attractive subjects for astrophysical studies for several reasons. Being in a transitional and therefore relatively short evolutionary phase, they often exhibit observable period changes associated with the shrinking of the orbit, as well as signs of weak mass accretion from the M dwarf (MD) to the white dwarf (WD). Since the MD is tidally locked to the WD, it rotates much more rapidly than what would be expected from a single main sequence star of the same mass and age. This in turn causes heightened stellar activity, which is indeed observed in almost all PCEBs regardless of spectral type of the main sequence component \citep{rebassa13}. 

Activity in M dwarfs is an interesting subject in its own right. As the oldest and most numerous main sequence stars in the Galaxy, M dwarfs are being studied as promising hosts for habitable extrasolar planets. Magnetic activity can affect the precision of radial velocity measurements and thus limit our ability to detect low-mass rocky planets in the habitable zone \citep{barnes15}. It might also be related to the known discrepancy between the radii predicted by stellar structure models and those derived from observations of eclipsing binaries with M dwarf components \citep{ribas06, morales10}. PCEBs are ideal for determination of stellar parameters because the eclipses of the WD are very sharp and allow for precise radius measurements that can be used to test and calibrate the models.

QS Vir has received a lot of attention since its discovery in the Edinburgh-Cape blue object survey \citep{stobie97} and the seminal work of \citet{donoghue03} because in addition to having all the characteristics of a detached pre-CV it also shows signs of significant and variable accretion \citep{matranga12}. \citet{donoghue03} initially classified QS Vir as a hibernating cataclysmic variable, but it has since been demonstrated in several studies that the MD, although close to filling its Roche lobe, doesn't yet actually fill it \citep{ribeiro10, parsons11, parsons16}. \citet{matranga12} examine several possible explanations for the observed strength and variations of accretion in the absence of Roche overflow, but find none that are entirely satisfactory.

The system also displays significant and seemingly cyclic variations in eclipse times, recently discussed by \citet{bours16}. The possibility that they are caused by Applegate's mechanism \citep{applegate92} was considered and rejected by \citet{parsons10} on the grounds that the energy required for variations with observed amplitude was by an order of magnitude larger than the energy output of the MD. \citet{almeida11} interpreted them as a result of perturbations by two circumbinary planets, but the proposed orbital configuration has been found to be unstable by \citet{horner13}, leaving the variations unexplained.

The most recent study of QS Vir (\citealt{parsons16}, hereafter \citetalias{parsons16}) presents an analysis of high resolution spectroscopy that, among other results, confirms significant spot coverage and a high level of magnetic activity of the MD component.

We made new multicolor observations of QS Vir with the 1.88 m telescope of the Kottamia Observatory in Egypt in 2015 and 2016, and analyzed these data using the binary system modeling software of G. Djura{\v s}evi{\' c} \citep{djur92a, djur98}. Initially, we tested models with two and three dark spots on the MD. These gave physical and orbital parameters of the system in good agreement with the findings of previous studies, but failed to produce satisfactory simultaneous fits to light curves observed in \textit{BVRI} filters. Individual fits were possible, but with diverging spot parameters (Section \ref{ssSimul}). To investigate the issue, we included in our analysis all CCD light curves available in online archives. Our final dataset samples almost 25 years through six observational seasons. But the problems regarding simultaneous fitting of light curves in multiple filters persisted with the archival data too.

A grid-search procedure was finally devised to examine the parameter space of spot coordinates, determine the optimal number of spots and consistent spot locations across different filters (Section \ref{ssCoords}). We find that a single dark spot is sufficient to account for all the asymmetries in the light curves (Section \ref{ssCount}), and that adding more spots does not improve the fit of the models to the observations. While this does not preclude the existence of additional spots on the MD, it shows that the light curves alone do not contain enough information for reliable modeling of more than one dominant feature. We further find that this dominant feature has been stable for the past 15 years, after a major migration from a different location that happened between observations made in 1993 and 2002 (Section \ref{sRes}). Given the high level of magnetic activity of the MD, we propose that this event was a part of a flip-flop cycle \citep{korhonen01}.


\section{New observations}
\label{sNObs}

Photometric observations of QS Vir in Bessell \textit{BVRI} filters were carried out on six nights in March 2015 and on another two nights in March and April 2016 with the EEV CCD 42-40 camera ($2048\times2048$ pixels) cooled by liquid nitrogen to -125\celsius\ and attached to the Newtonian focus of the 1.88 m reflector telescope of Kottamia Observatory in Egypt. A total of 786 CCD science frames (432 in 2015 and 354 in 2016) were obtained and reduced using the Muniwin software package\footnote{http://c-munipack.sourceforge.net}. Observations in the \textit{VRI} filters were done sequentially\footnote{Filters were changed after every observation in sequence.} in four observing sessions, and in the \textit{B} filter separately\footnote{Observations were taken in only one filter.} in another four sessions. The summary of observations, with starting and ending dates and times of observations on each night and the counts of science frames in each filter, is given in Table \ref{tObsSum}.

Differential photometry was performed with respect to UCAC4 384-063825 (comparison star) and UCAC4 385-063763 (check star). Identifiers, coordinates and B, V, J and H magnitudes from the UCAC4 catalog \citep{zach12} are given in Table \ref{tStars} for the variable, comparison and check stars. Exposure times were 300, 120, 60 and 30 seconds with the B, V, R and I filters, respectively. The long exposures in the B filter were necessary because of the declining condition of the mirrors.

The complete light curves are available in the online version of the article. An excerpt is given in Table \ref{tLC} for guidance regarding its form and content.


\begin{table}
\caption{Summary of new observations.}
\label{tObsSum}
\vspace{10pt}
\centering
\footnotesize{
\begin{tabular}{lccrrrrr}
\tableline\tableline 
HJD		&	Start Date and Time	&	End Date and Time	&	B	&	V	&	R	&	I	&	Total	\\
\tableline
\noalign{\smallskip}
2457101	&	19-Mar-15 23:02:24	& 	20-Mar-15 02:47:02	& 		& 	46	& 	47	& 	47	& 	140	\\
2457102	&	20-Mar-15 22:55:12	& 	21-Mar-15 02:40:10	& 		& 	51	& 	48	& 	44	& 	143	\\
2457104	&	23-Mar-15 00:05:49	& 	23-Mar-15 02:46:30	& 	22	& 		& 		& 		& 	22	\\
2457105	&	23-Mar-15 22:19:35	& 	24-Mar-15 03:05:17	& 	42	& 		& 		& 		& 	42	\\
2457106	&	24-Mar-15 23:38:24	& 	25-Mar-15 02:44:03	& 	32	& 		& 		& 		& 	32	\\
2457107	&	26-Mar-15 00:27:55	& 	26-Mar-15 01:51:57	& 		& 	17	& 	19	& 	17	& 	53	\\
2457479	&	31-Mar-16 21:58:44	& 	01-Apr-16 02:46:44	& 	93	& 		& 		& 		& 	93	\\
2457482	&	03-Apr-16 21:15:13	& 	04-Apr-16 02:12:29	& 		& 	86	& 	89	& 	86	& 	261	\\
\tableline
Total	& 						&						&	189 &	200 &	203 &	194 &	786	\\
\tableline
\end{tabular}}
\tablecomments{The B, V, R and I columns contain the number of observations in the corresponding filter 
taken on the corresponding night. The Total column shows the number of observations across all filters taken on the corresponding night. The Total row shows the number of observations taken in the corresponding filter across all nights.}
\end{table}


\begin{table}
\caption{Information about the variable, comparison and check stars.}
\label{tStars}
\vspace{10pt}
\centering
\footnotesize{
\begin{tabular}{lccccccc}
\tableline\tableline 
Star		& UCAC4 ID			& $\alpha_{2000}$ 	& $\delta_{2000}$ 	& B 		& V 		& J 		& H 	\\
\tableline
\noalign{\smallskip}
QS Vir 		& UCAC4 384-063823	& 13:49:51.950 		& -13:13:37.50 		& 14.984 	& 14.400 	& 10.829 	& 10.271\\
Comparison	& UCAC4 384-063825	& 13:49:58.146 		& -13:13:58.88 		& 15.442 	& 14.867 	& 13.716 	& 13.427\\
Check		& UCAC4 385-063763	& 13:49:51.888		& -13:10:58.87		& 14.176	& 13.415	& 12.000 	& 11.602\\
\tableline
\end{tabular}}
\end{table}


\begin{table}
\caption{New BVRI light curves of QS Vir.}
\vspace{10pt}
\label{tLC}
\centering
\begin{tabular}{rrrcc}
\tableline\tableline
HJD			&	Phase	&	Magnitude	&	Filter		&	Season	\\
\tableline
\noalign{\smallskip}
2457104.504	&	0.21739	&	14.99045	&	B			&	S15		\\
2457104.508	&	0.24678	&	14.97470	&	B			&	S15		\\
2457104.523	&	0.34475	&	14.95250	&	B			&	S15		\\
2457104.534	&	0.41858	&	14.99359	&	B			&	S15		\\
2457104.538	&	0.44305	&	15.00112	&	B			&	S15		\\
2457104.542	&	0.46780	&	14.99084	&	B			&	S15		\\
2457104.545	&	0.49221	&	15.01275	&	B			&	S15		\\
2457104.549	&	0.51662	&	14.99965	&	B			&	S15		\\
2457104.553	&	0.54109	&	14.99819	&	B			&	S15		\\
2457104.557	&	0.56550	&	14.99046	&	B			&	S15		\\
\tableline
\end{tabular}
\tablecomments{The complete light curves of QS Vir obtained in 2015 and 2016 with the 1.88 m telescope of Kottamia Observatory in Egypt, using the Bessell \textit{BVRI} filters. Table \ref{tLC} is available online at [CDS address]. Only an excerpt is shown here for guidance regarding its form and content.}
\end{table}


\subsection{Ephemeris and times of minimum light}
\label{ssEph}

Twenty-two eclipses were recorded in different filters during our observations. The exact times of minimum light were determined using the software package AVE \citep{barbera}, which is based on the method of \citet{kwandvw56}. Using these measurements and the orbital period taken from \citet{parsons10}, we updated the ephemeris (Equation \ref{eqEph}) with the software package Peranso\footnote{http://www.peranso.com}. The final linear ephemeris that we use with all data sets is given by Equation \ref{eqEph}:

\begin{equation}
\label{eqEph}
HJD (Min I) = 2457102.511412 + 0.1507575 \times E
\end{equation}

A comparison with the results of a recently published, long-term eclipse timings study done by \citet{bours16}, based on exquisite, high-speed photometry, shows that our own measurements of eclipse times are of too low precision to help constrain the nature of the period variation of QS Vir. The measurement errors of our eclipse timings are of the same order of magnitude as the amplitude of the period variation since we only have a few observations per minimum. They are nevertheless listed in Table \ref{tMin}. 


\begin{table}
\caption{Eclipses recorded during our observations.}
\vspace{10pt}
\label{tMin}
\centering
\begin{tabular}{rccrr}
\tableline\tableline
Date			& Type\tablenotemark{a}	
				& Filter	
				& Time\tablenotemark{b}	
				& Error													\\[-3mm]
				&		&			&	(HJD)				&			\\
\tableline
\noalign{\smallskip}
20 Mar 2015 	&	I	&	V		&	2457101.60684		&	0.00216	\\
20 Mar 2015 	&	I	&	R		&	2457101.60806		&	0.00216	\\
20 Mar 2015 	&	I	&	I		&	2457101.60573		&	0.00215	\\
21 Mar 2015 	&	I	&	V		&	2457102.51241		&	0.00218	\\
21 Mar 2015 	&	I	&	R		&	2457102.51058		&	0.00216	\\
21 Mar 2015 	&	I	&	I		&	2457102.51129		&	0.00218	\\
21 Mar 2015 	&	II	&	V		&	2457102.58619		&	0.00216	\\
21 Mar 2015 	&	II	&	R		&	2457102.58743		&	0.00217	\\
21 Mar 2015 	&	II	&	I		&	2457102.58816		&	0.00219	\\
25 Mar 2015 	&	II	&	B		&	2457106.50664		&	0.02551	\\
25 Mar 2015 	&	I	&	B		&	2457106.58512		&	0.00509	\\
26 Mar 2015 	&	II	&	V		&	2457107.56280		&	0.00215	\\
26 Mar 2015 	&	II	&	I		&	2457107.56321		&	0.01143	\\
2 Apr 2016		&	II	&	V		&	2457481.43922		&	0.00136	\\
2 Apr 2016		&	II	&	R		&	2457481.44020		&	0.00046	\\
2 Apr 2016 		&	II	&	I		&	2457481.44020		&	0.00034	\\
3 Apr 2016 		&	II	&	V		&	2457482.49604		&	0.00033	\\
3 Apr 2016 		&	II	&	R		&	2457482.49470		&	0.00024	\\
3 Apr 2016 		&	II	&	I		&	2457482.49570		&	0.00043	\\
3 Apr 2016 		&	I	&	V		&	2457482.56910		&	0.00236	\\
3 Apr 2016 		&	I	&	R		&	2457482.56990		&	0.00237	\\
3 Apr 2016 		&	I	&	I		&	2457482.57060		&	0.00237	\\
\tableline
\end{tabular}
\tablenotetext{a}{Type of eclipse: "I" stands for the primary (deeper), and "II" for the secondary (shallower) eclipse.}
\tablenotetext{b}{Time of eclipse, derived from the observations as described in Subsection \ref{ssEph}.}
\end{table}


\section{Archival observations}
\label{sAObs}

Initially we intended to analyze only the newly observed light curves, since previous studies of QS Vir were quite thorough. But when we encountered issues with simultaneous fitting of the model to light curves in all filters (to be addressed in Section \ref{ssSimul}), we committed to the analysis of all available CCD photometry. This includes:

\begin{itemize}

\item{Johnson-Cousins \textit{VRI} light curves observed sequentially on four nights in June 1993 with the SAAO 1-m telescope using the UCL camera with the RCA chip (hereafter season S93), analyzed and published by \citet{donoghue03}. There was a flare event on the MD during one of the nights; observations affected by it were easy to identify and we removed them.}

\item{Johnson \textit{BR} light curves recorded separately during three nights in April 2002 with the Mt Stromlo Observatory 74-inch telescope, the Monash Imager and a 2K $\times$ 4K camera (hereafter season S02K), analyzed and published by \citet{kawka02}.}

\item{Sloan \textit{u'g'r'} light curves recorded simultaneously\footnote{Observations were made in all filters at the same time.} on May 20, 2002 with the 4.2-m William Herschel Telescope on La Palma (hereafter season S02P), analyzed and published by \citet{parsons10}; and Sloan \textit{u'g'i'} light curves recorded simultaneously on April 21, 2010 with the 3.5-m New Technology Telescope on La Silla (hereafter season S10), analyzed and published by \citetalias{parsons16}. Both sets comprise high-speed photometry obtained with ULTRACAM \citep{dhillon07}. QS Vir was observed by the same group and with similar equipment in 2003, 2006 and 2011 too, but these light curves are incomplete (mostly focusing on the primary minimum) and are not suitable for seasonal modeling.}

\end{itemize}

All the light curves, including our own observations from 2015 and 2016 (hereafter seasons S15 and S16, respectively) were folded to phases according to the ephemeris given in Equation \ref{eqEph} and normalized to the higher maximum (phase 0.75 for S93 and phase 0.25 for all other seasons). They are shown together with the models in Figs. \ref{fS93} through \ref{fS16}.


\section{Modeling of the light curves}
\label{sMod}

The first step in our study of QS Vir was exploratory modeling of the new observations (seasons S15 and S16) using the program by \citet{djur92a} generalized for the case of contact configurations \citep{djur98} and updated to make use of the limb-darkening coefficients for white dwarfs from \citet{giann13}. The program implements a robust binary star model based on Roche geometry that can be applied on a wide variety of binary configurations, including those with an accretion disk \citep[e.g.][]{mennickent12}. A model of QS Vir with an accretion disk around the WD was tested in this preliminary phase and we found that photometric data does not support it. We instead adopted a marginally detached configuration with the MD just barely underfilling its critical Roche surface. The apparent asymmetry of the light curves can be reproduced by placing one or more dark spots on the MD.

A comprehensive list of model parameters, describing all the major physical processes in close binaries, can be found in our previous publications \citep[see e.g.][]{caliskan2014}. For the present work, we adopted the spectroscopic elements (the mass ratio, $q$, and the orbital separation, $a_{\rm orb}$) as well as the effective temperature of the WD ($T_{\rm WD}$) from \citetalias{parsons16} and kept them constant. The reflection coefficients ($A_{\rm WD}$ and $A_{\rm MD}$) and gravity darkening exponents ($\beta_{\rm WD}$ and $\beta_{\rm MD}$ ) were also kept fixed to the theoretical values appropriate for each component \citep{vonz,lucy,rucinski69}. We assume the components are tidally locked and rotate synchronously with the orbital motion, so that the nonsynchronous rotation coefficient, $f=\omega_{rot}/\omega_{orb}$, is constant and equal to 1. The contrast of the dark spots ($C$), defined as the ratio of the effective temperature of an affected elementary surface with and without the spot, is also kept constant at a value of 0.9. This roughly corresponds to a temperature difference of $\Delta \rm{T= 300K}$ which is appropriate for a spotted M dwarf star according to \citet{berdyugina05}. A summary of fixed parameters is given in Table \ref{tFixed}.


\begin{table}
\caption{Orbital and stellar parameters adopted from literature.}
\vspace{10pt}
\label{tFixed}
\centering
\begin{tabular}{lr} 
\tableline\tableline
Parameter						&  Value		\\
\tableline
\noalign{\smallskip}
$q=m_{\rm MD}/m_{\rm WD}$		& 0.489			\\
$a_{\rm orb}\ [R_{\odot}]$		& 1.253			\\
$P\ [d]$						& 0.1507575 	\\
$T_{\rm WD}\ [K]$				& 14200 		\\
$A_{\rm WD}$					& 1.0			\\
$A_{\rm MD}$					& 0.5			\\
$\beta_{\rm WD}$				& 0.25			\\
$\beta_{\rm MD}$				& 0.08			\\
$f_{\rm WD}=f_{\rm MD}$			& 1.0			\\
${C=T_{\rm Spot}/T_{\rm MD}}$	& 0.9			\\
\tableline
\end{tabular}
\tablecomments{These parameters were kept fixed to the listed values in all stages of modeling. Here and in the entire paper, subscripts WD and MD refer to the white dwarf and the M dwarf, respectively.}
\end{table}

Treatment of limb darkening follows the nonlinear approximation of \citet{claret00}, with the coefficients for the appropriate filters interpolated from their tables based on the current values of effective temperature and effective gravity in each iteration. Limb-darkening coefficients for the WD were taken from \citet{giann13}. The reflection effect is accounted for by applying a temperature correction to affected elementary surfaces according to the prescription of \citet{khruzina1985}.

Spots are modeled as circular regions of uniform fillout and constant temperature contrast. This is clearly a rough approximation of what actually goes on on the surfaces of active stars \citep[see e.g.][]{parsons16, barnes15}. However, its usefulness has been proven in numerous studies of binary stars of all spectral types, including M dwarfs. \citet{wilson17} recently published a study of CU Can, one of the few known eclipsing binaries in which both components are M dwarfs. Atmospheric activity, apparent in the asymmetry of the light curves, is explained with two dark, low-contrast spots on top of a binary system model based on Roche geometry similar to the one used in the present study. The same approach was taken in the study of another M dwarf binary, BX Tri, by \citet{dimitrov10}. This system is more similar to QS Vir, as it has a very short period of $P\approx 0.2 d$ and one of the components nearly fills its Roche lobe. As we go on to show in the following sections, the light curves of QS Vir can be adequately modeled with a single dark spot on the MD, likely representing a group of smaller spots similar to Doppler images presented by \citet{barnes15}.

\subsection{Simultaneous and individual fitting of filter-specific light curves}
\label{ssSimul}

Initial photometric solutions, comprising the orbital and stellar parameters that produce a model which optimally fits the new observations, were found using the Marquart-Levenberg algorithm \citep{marquardt} with modifications described in detail in \citet{djur92b} to minimize the sum of squared residuals between the observed and calculated light curves. The following parameters were adjusted:  the effective temperature of the MD ($T_{\rm MD}$), the orbital inclination ($i$), the filling factor, defined as the ratio of the polar radius and the critical polar radius for each component ($F_{\rm WD}$ and $F_{\rm MD}$); and for each dark spot on the MD: the longitude ($\lambda$), measured clock-wise from the intersection of the line of centers and the back of the MD (opposite to the $L_1$ Lagrange point) with values from 0$^{\circ}$ to 360$^{\circ}$, the latitude ($\varphi$) measured from the stellar equator towards the poles with values from -90$^{\circ}$ to 90$^{\circ}$, and the angular radius ($\theta$).

The standard approach in the analysis of multicolor light curves of eclipsing binaries is to fit the model simultaneously to observations in all filters, but in the case of QS Vir this resulted in fits of poor quality for both S15 and S16. Modeling each light curve individually produced fine fits to the data, but resulted in different parameters for each filter. The differences in main binary and stellar parameters (orbital inclination, temperature of the MD and the sizes of the stars) were within expected uncertainties; however, the coordinates and sizes of the dark spots on the MD varied significantly from filter to filter.

To eliminate the possibility that the issue arose from some fault with our data, we then analyzed the archival light curves (seasons S93, S02K, S02P and S10). In each of these seasons, simultaneous fitting of all filter failed, and individual fits resulted in significantly different spot configurations, same as with seasons S15 and S16.

Assuming the spots are magnetic in origin, such variations may be due to different configurations of the magnetic field at different depths in the atmosphere of the MD that we observe in different filters. Another possibility is the presence of additional magnetic structures, such as plages, in the vicinity of the spots. These phenomena might explain small variations in spot positions and moderate variations in spot sizes. However, the variations we found were significant. We judged these results to be implausible and attempted a different approach to finding the optimal photometric solution.

\subsection{Search for consistent spot coordinates}
\label{ssCoords}

In the following sections, we consider the question: can a model be found with consistent (or fixed) spot locations, that fits all individual light curves within an observational season, if we allow for minor variations in other model parameters (including spot size)? 

We try to answer it by creating a coordinate grid covering the surface of the MD, and fitting a model to each light curve of a given season at each node of the grid, keeping spot coordinates fixed to grid values. We use the $\alpha$-constrained Nelder-Mead Simplex \citep{takahama03} for model optimization at grid nodes to avoid certain hard-coded behaviors of the Marquart-Levenberg algorithm implemented by \citet{djur92b}, such as unconditional fitting of spot latitudes. 

This allows us to perform a survey of the parameter space of spot coordinates. For each light curve within a season, the procedure results in a list of trial solutions (one for each node in the grid) that can be ordered by the quality of the achieved fit (the reduced $\chi^2$ value). We then combine the filter-specific solutions in each node into a single "seasonal solution". 

A simple sum of filter-specific $\chi^2$ values is not the best measure for the quality of the seasonal solution because the worst fit will have the greatest weight in the sum. The worst fits are typically obtained for light curves in blue filters, dominated by the radiation from the WD, where the influence of spots on the MD is negligible. To get around this, we normalize the $\chi^2$ value of each trial model to the best achieved within a filter-specific trial set and use the normalized measure $\chi^2_n$ (Equation \ref{eNChi2}) to compare seasonal solutions.

\begin{equation}
\label{eNChi2}
\chi^2_n= \frac{\chi^2}{min(\chi^2)}
\end{equation}

The best individual solution for each light curve will have $\chi^2_n = 1$, and ideally the seasonal solution should have $\chi^2_n$ equal to the number of light curves within the season. But as expected after the preliminary analysis, different light curves within a season prefer different spot coordinates so the $\chi^2_n$ values for the best seasonal solutions are larger than that (see Table \ref{tRes}).

Spot coordinates chosen in this way are usually found among the top 10 solutions for each light curve (with $\chi^2$ values roughly up to 10\% larger than the best) and produce synthetic light curves that are negligibly different from filter-specific optimal solutions (see Figs. \ref{fS93} to \ref{fS16}).
 
\subsection{One or more spots?}
\label{ssCount}

\citetalias{parsons16} identified three dark spot-like regions on the MD, but our preliminary analysis indicated that the light curves in all six seasons can be adequately modeled with only one spot (see discussion in Section \ref{ssComp}). Since the search procedure we developed requires calculation of trial models in numbers that increase exponentially with dimensions and precision of the grid, we first constructed low-resolution grids with a step of $30^{\circ}$ in both longitudes and latitudes for one-spot and two-spot models to determine the merit of including additional spot(s). 

Unsurprisingly, solutions obtained for two-spot models turned out to be statistically superior to those for one-spot models, but visual inspection of the best solutions from one- and two-spot grids (Fig. \ref{fComp}) confirmed the results of the preliminary analysis and convinced us that the inclusion of a second (or third) spot does not significantly improve the fit of the model to data. 

This result doesn't mean that there's only one spot on the MD component of QS Vir. It only shows that the light curves available to us do not contain enough information to uniquely identify and parameterize more than one spot. Thus, we limit the analysis to models with one spot.

\subsection{Five degree grid for one spot}
\label{ssGrid5}

The choice of precision (grid step) in spot coordinates was informed by comparing the quality of filter-specific and seasonal solutions for test grids with steps of 30$^{\circ}$,  10$^{\circ}$, 5$^{\circ}$ and 2$^{\circ}$. These test grids were computed for only one season (S16) because the computation of 2$^{\circ}$ grids is quite time-consuming (of the order of 10 days with the available equipment\footnote{8 $\times$ Intel Core i7-4770K CPU with 16 GB RAM.}). The comparison showed that switching from 30$^{\circ}$ to 10$^{\circ}$ grid brought the greatest improvement in quality of the solutions, with clear signs of diminishing returns for increasing precision. The solutions for 5$^{\circ}$ and 2$^{\circ}$ grids were almost indistinguishable. We therefore perform the complete analysis with 5$^{\circ}$ grids. 

Principal physical parameters of the system and its components that are adjusted at every node of the spot coordinate grid are summarized in Table \ref{tAdjust}. Initial values of the effective temperature for the MD, orbital inclination and the filling factors were adopted from \citetalias{parsons16}; the initial value for spot size is close to the mean spot size over all filters and seasons obtained in preliminary analysis. Two additional parameters, the phase shift and the magnitude shift, were also fitted for small (of the order of $10^{-5} - 10^{-3}$) translations along the phase and magnitude axes.


\begin{table}
\caption{Model parameters adjusted for each light curve by grid-search or optimization.}
\vspace{10pt}
\label{tAdjust}
\begin{tabular*}
{\textwidth}{l@{\extracolsep{\fill}}lrrl@{\extracolsep{\fill}}rrrrr} 
\tableline\tableline
 & 	& \multicolumn{2}{c}{Fixed to grid values} &	 & \multicolumn{5}{c}{Optimized} 		\\
\cline{3-4}
\cline{6-10}
\noalign{\smallskip}
Parameters 		& & $\lambda\ [^{\circ}]$ & $\varphi\ [^{\circ}]$ & & $\theta\ [^{\circ}]$ 
				& $i\ [^{\circ}]$ & $T_{\rm MD}\ {\rm [K]}$ & $F_{\rm MD}$ & $F_{\rm WD}$	\\
\noalign{\smallskip}
Initial value	&	& -		& -		&	& 40	& 77.7	& 3100	& 0.977	& 0.020				\\
From			&	& 0		& -90	&	& 10	& 75	& 3000	& 0.950	& 0.019				\\
To				&	& 355	& 90	&	& 90	& 80	& 3500	& 0.990	& 0.022				\\
Step			&	& 5		& 5		&	& -		& -		& -		& -		& -					\\
\tableline
\end{tabular*}
\end{table}


\section{Results and discussion}
\label{sRes}

\begin{figure}
\caption{Data (black circles), filter-specific fits (orange line) and seasonal fit (green line) with corresponding O-C residuals for season S93.}
\label{fS93}
\medskip
\centering
\includegraphics[width=0.5\textwidth]{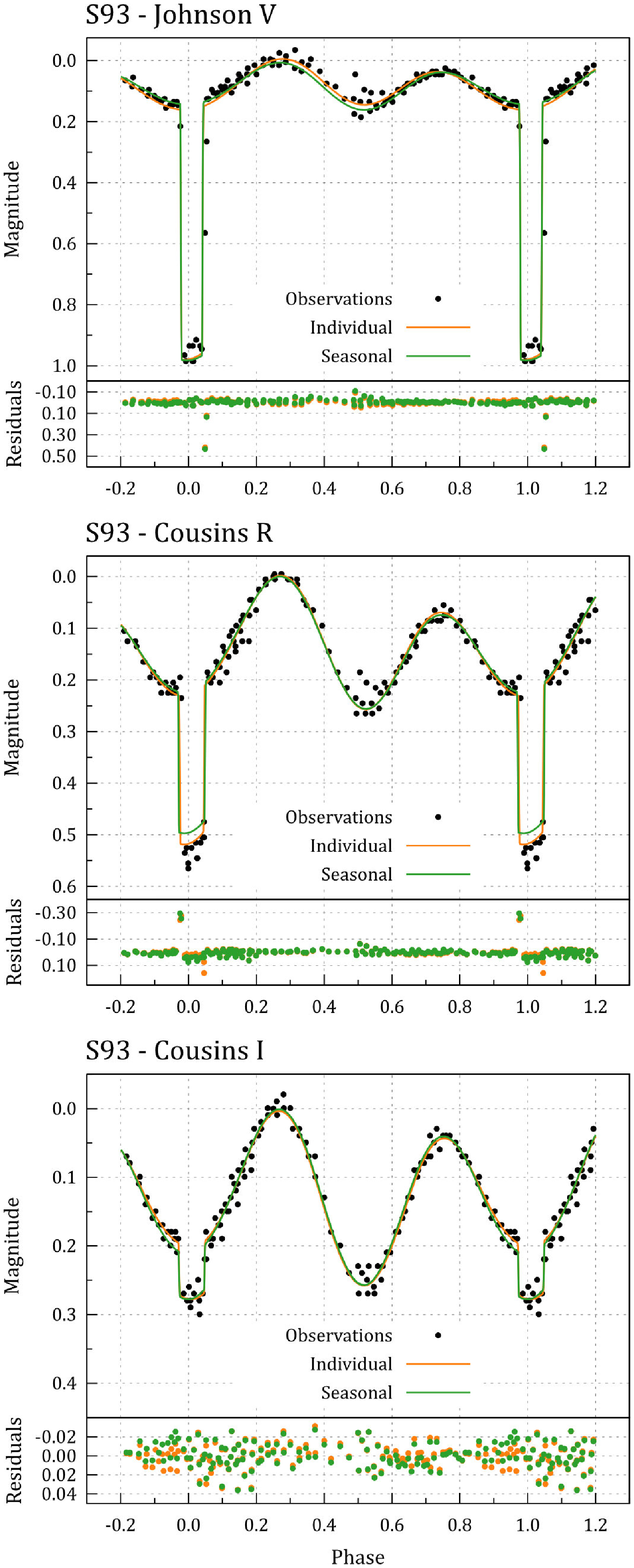}
\end{figure}

\begin{figure}
\caption{Data (black circles), filter-specific fits (orange line) and seasonal fit (green line) with corresponding O-C residuals for season S02K.}
\label{fS02K}
\medskip
\centering
\includegraphics[width=0.5\textwidth]{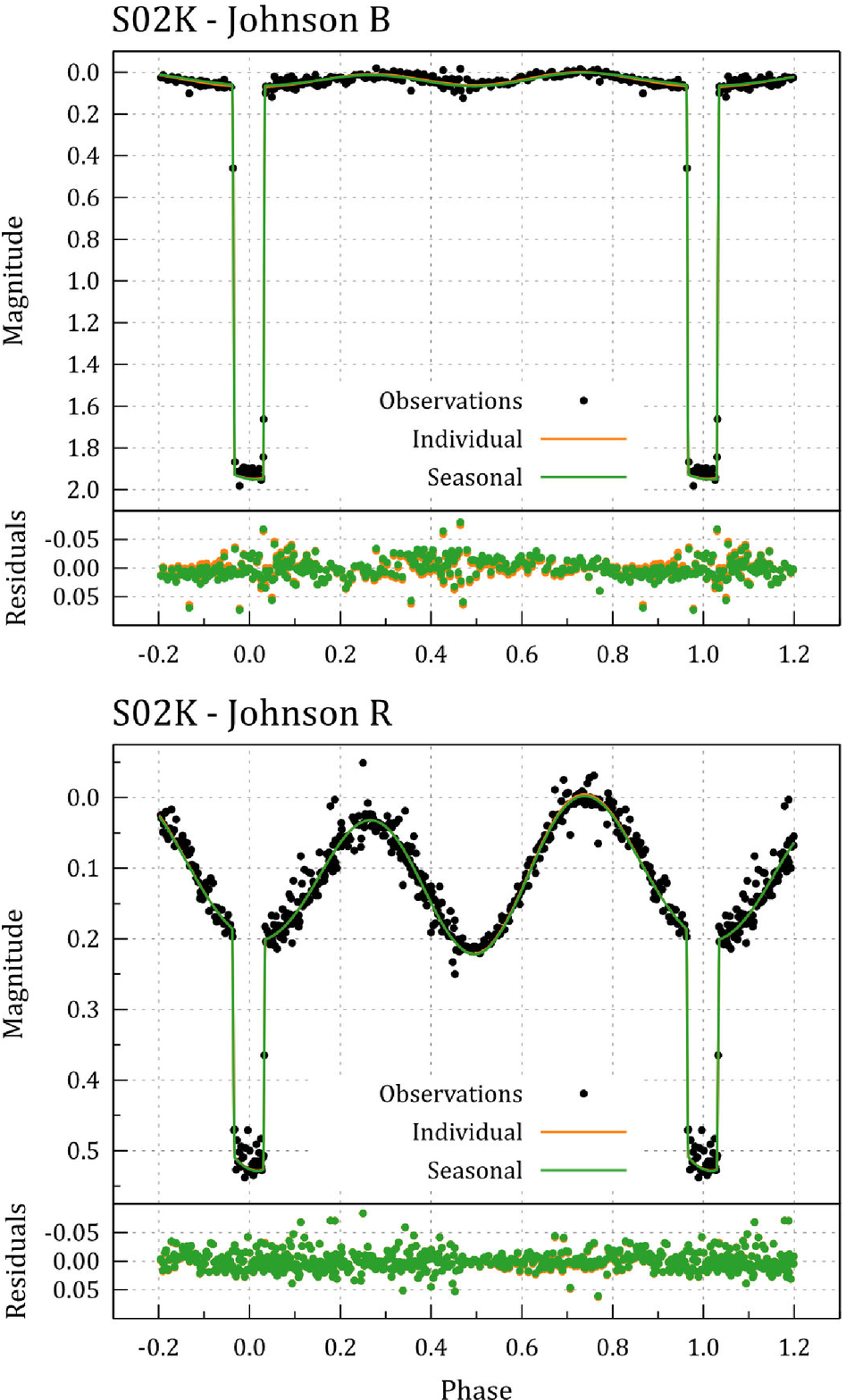}
\end{figure}

\begin{figure}
\caption{Data (black circles), filter-specific fits (orange line) and seasonal fit (green line) with corresponding O-C residuals for season S02P.}
\label{fS02P}
\medskip
\centering
\includegraphics[width=0.5\textwidth]{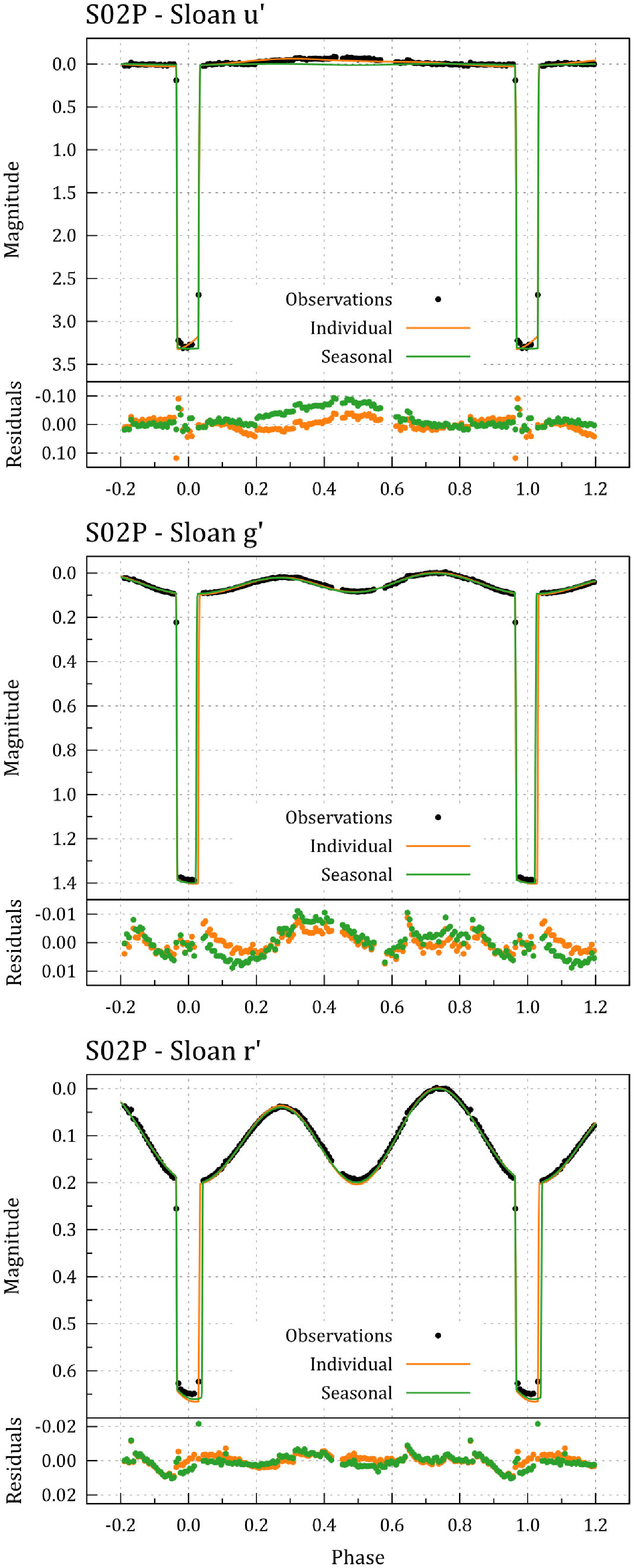}
\end{figure}

\begin{figure}
\caption{Data (black circles), filter-specific fits (orange line) and seasonal fit (green line) with corresponding O-C residuals for season S10.}
\label{fS10}
\medskip
\centering
\includegraphics[width=0.5\textwidth]{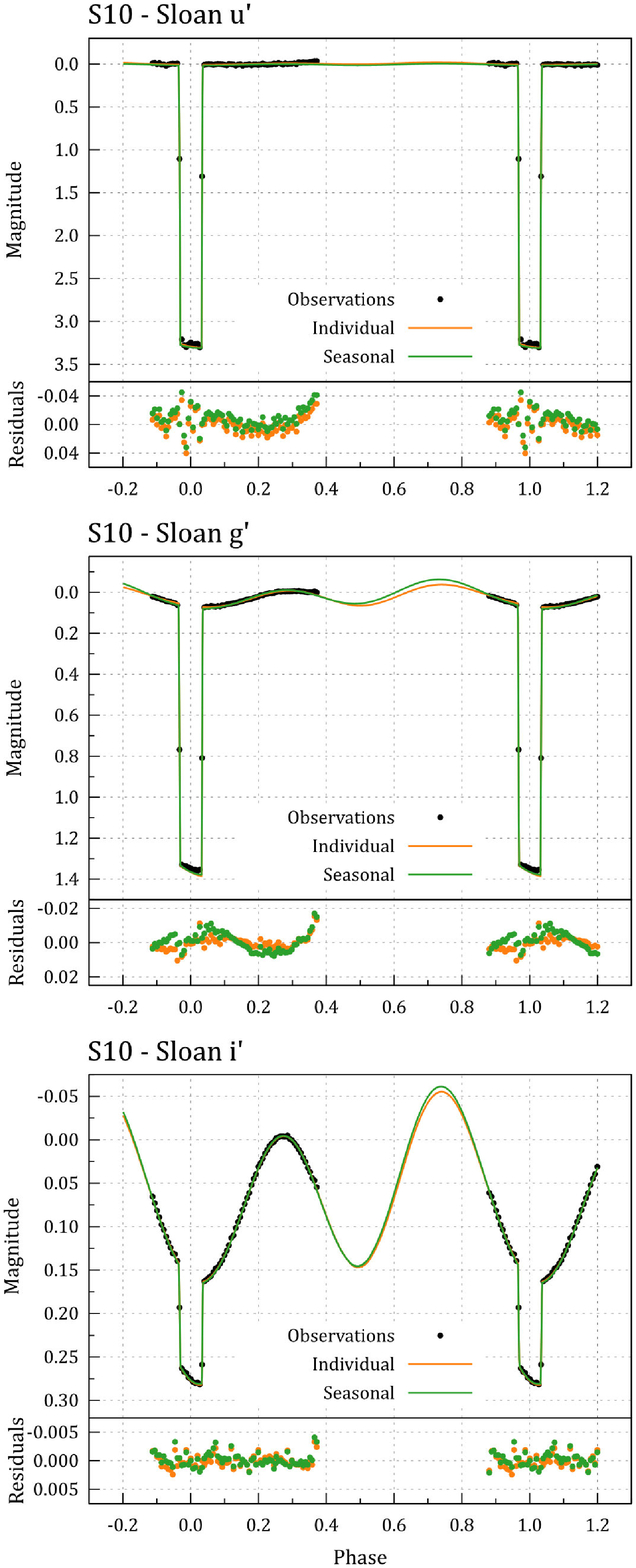}
\end{figure}

\begin{figure}
\caption{Data (black circles), filter-specific fits (orange line) and seasonal fit (green line) with corresponding O-C residuals season S15.}
\label{fS15}
\medskip
\centering
\includegraphics[width=\textwidth]{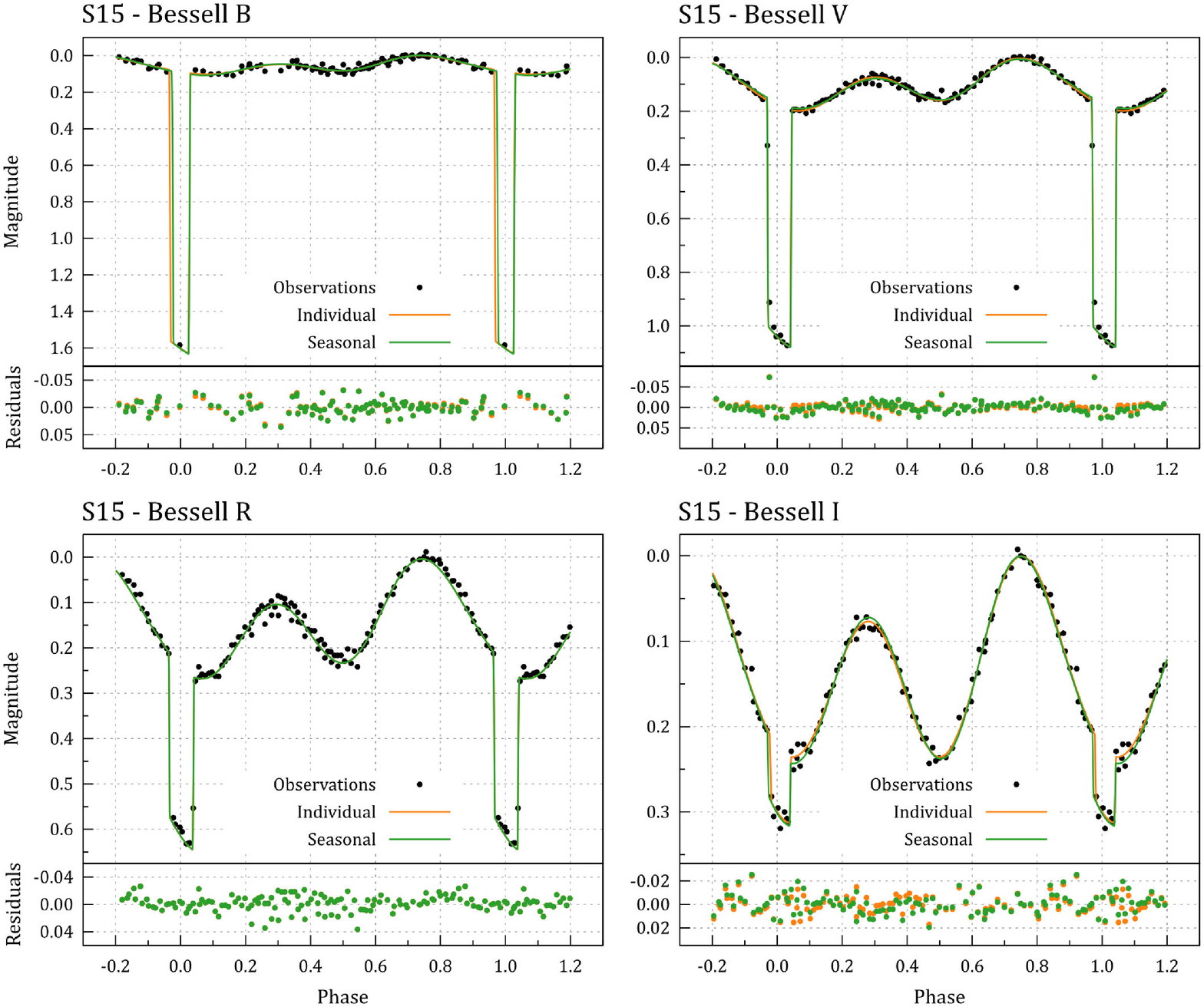}
\end{figure}

\begin{figure}
\caption{Data (black circles), filter-specific fits (orange line) and seasonal fit (green line) with corresponding O-C residuals for season S16.}
\label{fS16}
\medskip
\centering
\includegraphics[width=\textwidth]{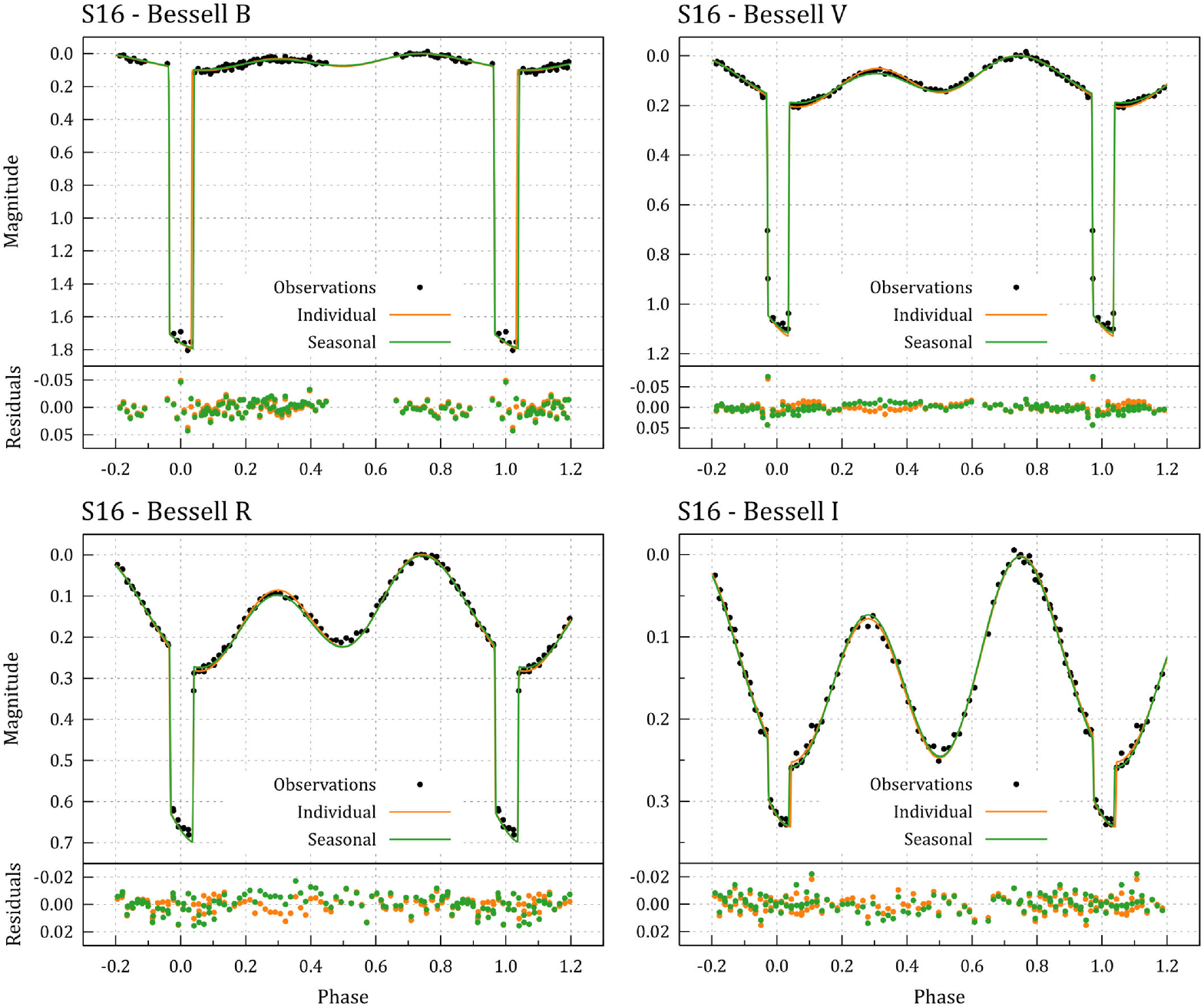}
\end{figure}

\begin{figure}
\caption{Comparison of best one-spot and two-spot models for Season S16. The two-spot models (blue lines) were calculated for the 30$^\circ$ grid only. One-spot models are shown for both the 30$^\circ$ grid (orange lines) and the final 5$^\circ$ grid (green lines).}
\label{fComp}
\medskip
\centering
\includegraphics[width=\textwidth]{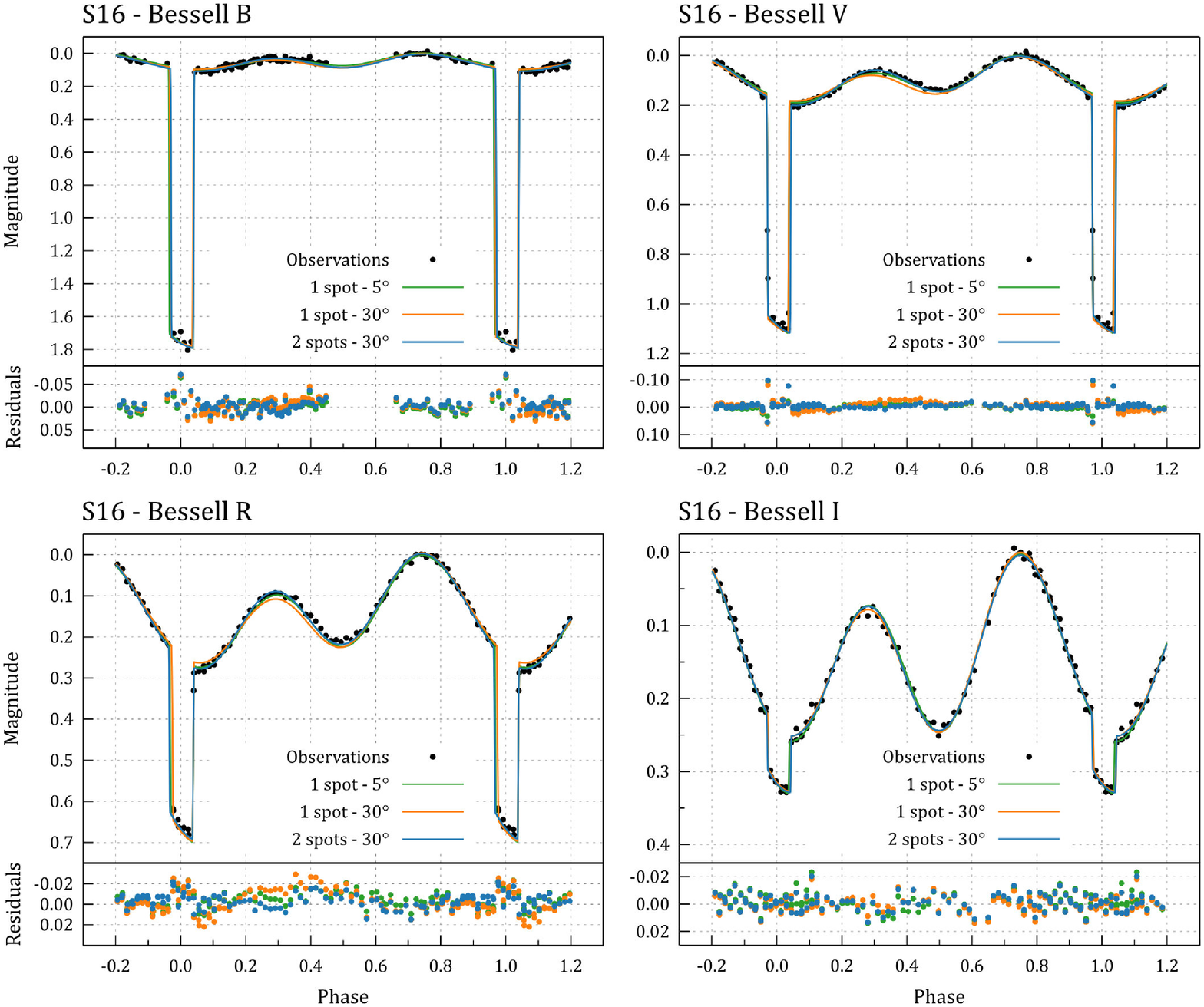}
\end{figure}

The final seasonal synthetic light curves are shown together with the observations and with individual, filter-specific fits for each of the six seasons in Figs. \ref{fS93} through \ref{fS16}. There is little to no deviation of seasonal fits from filter-specific fits. As a reminder, the seasonal fits all share the same spot model (the spot is located at the same longitude and latitude) for all the filters; whereas the filter-specific fits were made with spot coordinates as free parameters and resulted in models with spots located in different places in each filter. 


The differences are most notable in seasons S02P and S10 (Figs. \ref{fS02P} and \ref{fS10}, respectively), where seasonal fits are noticeably worse than filter-specific ones and there clearly exists a structure in the residuals. This structure resembles a cyclic variation that might be associated with additional spots, but adding a second, and then a third spot to the models failed to eliminate it. Note also that the structure in the residuals is almost invisible in the $r'$ and $i'$ filters, as are the differences between the seasonal and filter-specific fits, indicating that the discrepancies are related to something other than spots on the MD. Perhaps gaseous structures inside the critical Roche lobe of the WD found by \citet{parsons10} left an imprint on the light curves during these particular observations. 

Overall, the seasonal solutions fit the observations nearly as well as the filter-specific ones, which validates our assumption that a unique single-spot model that fits all the light curves within a given season with little to no variation in orbital and stellar parameters can indeed be found.

The chief results of our analysis are the longitudes of the dominant dark spot on the MD component of QS Vir (Table \ref{tRes}) over six observational seasons, sampling the period of almost 25 years (from 1993 to 2016). We find that it has been a stable feature for the past 15 years, with only minor variations in longitude. A major migration happened between season S93 and seasons S02K-S02P; the longitude of the spot in season S93 is nearly a mirror image of its longitude in all other seasons with regards to the plane defined by the poles of the MD and the $L_1$ Lagrange point (Fig. \ref{fSpots}).

\begin{figure}
\caption{$\chi_n^2$ values for best fitting models on each spot longitude present in the grid.}
\label{fLon}
\medskip
\includegraphics[width=\textwidth]{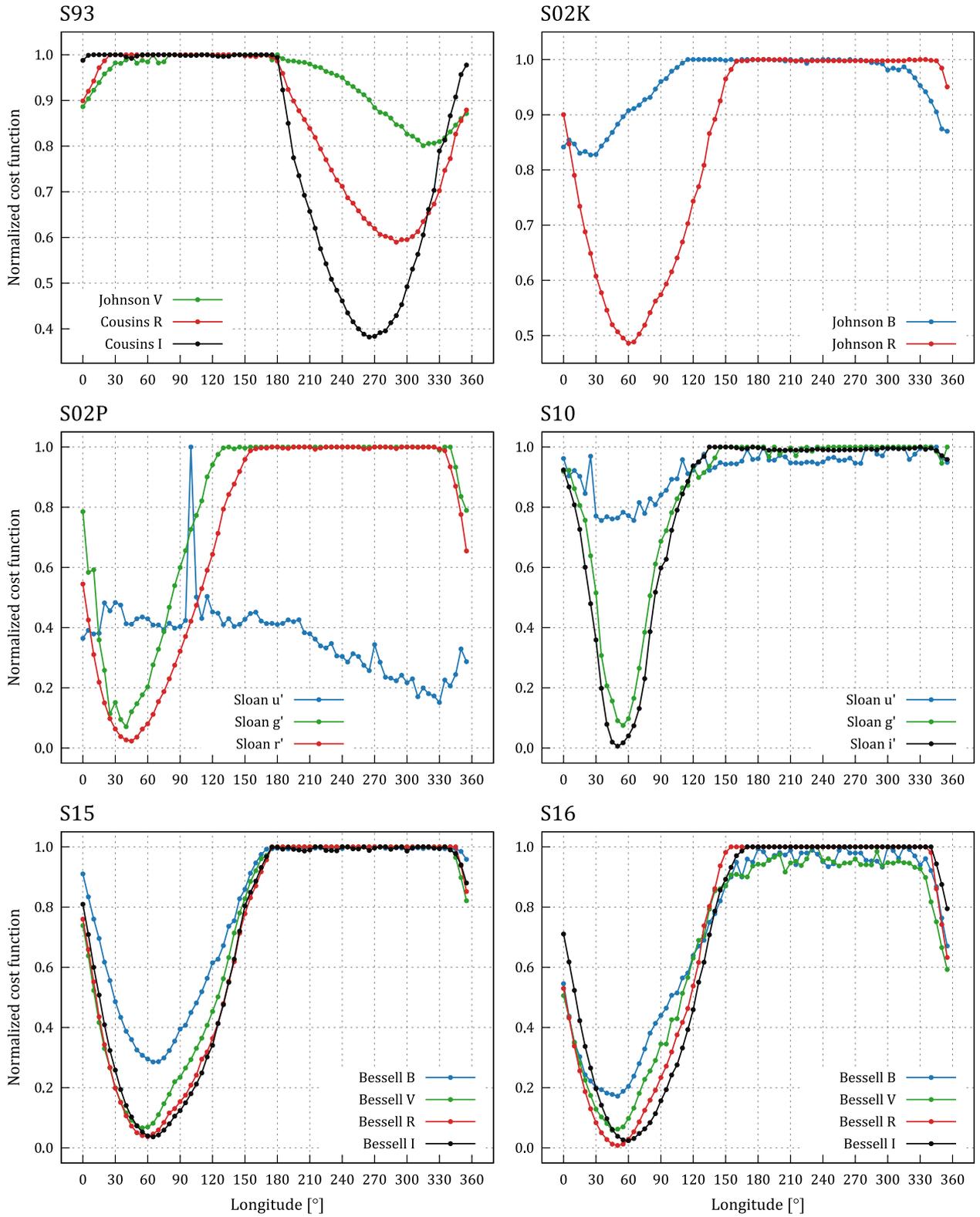}
\end{figure}

\begin{figure}
\caption{Goodness of fit (color) vs. spot longitude (on the x-axis) and spot latitude (on the y-axis) for all the light curves. Darker colors correspond to better fits. Every pixel is one node of the spot coordinate grid. White lines indicate active longitudes for seasonal solutions.}
\label{fContours}
\medskip
\includegraphics[width=\textwidth]{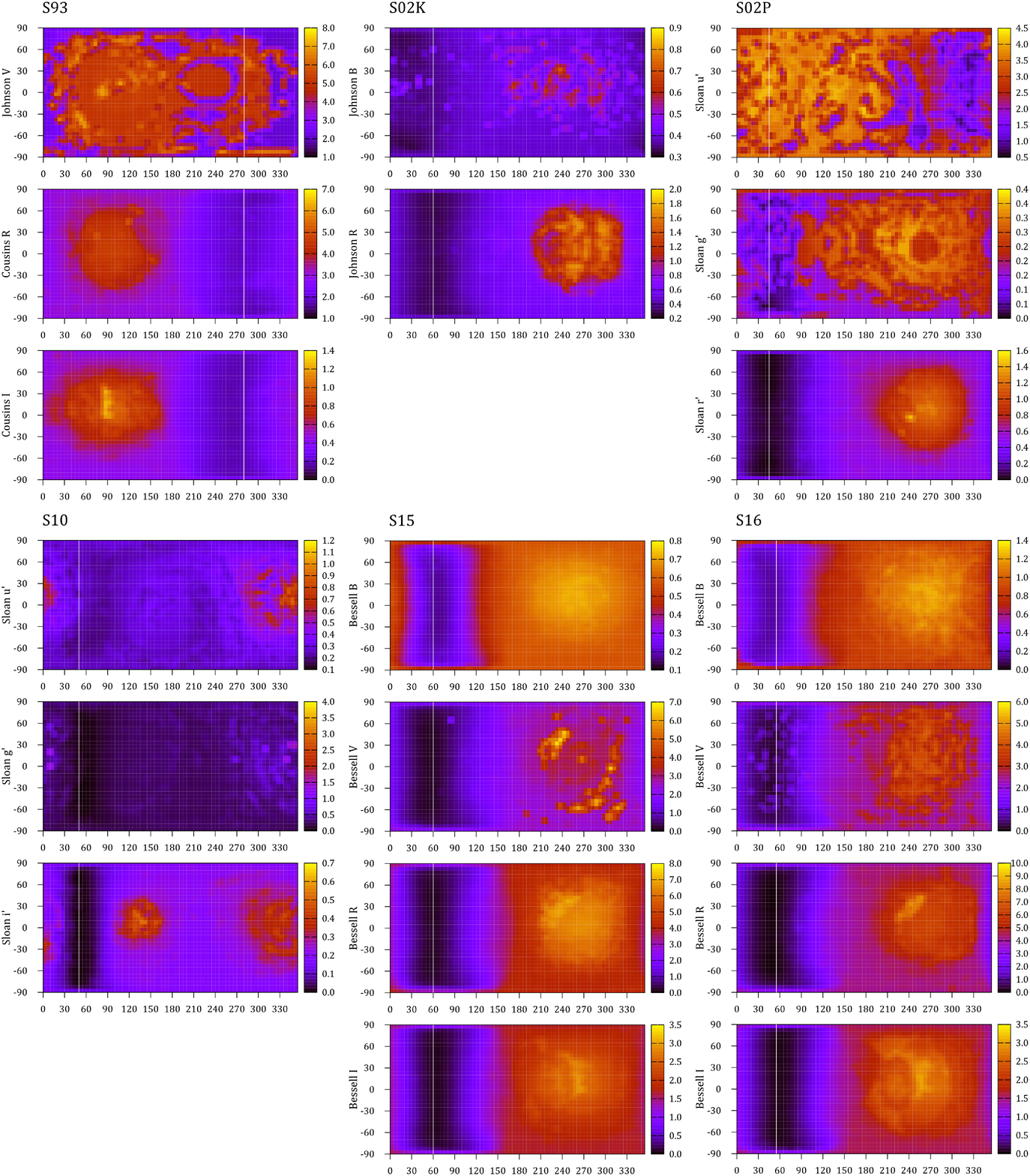}
\end{figure}

\begin{figure}
\caption{Appearance of QS Vir with the spot on the MD in different seasons.}
\label{fSpots}
\medskip
\includegraphics[width=\textwidth]{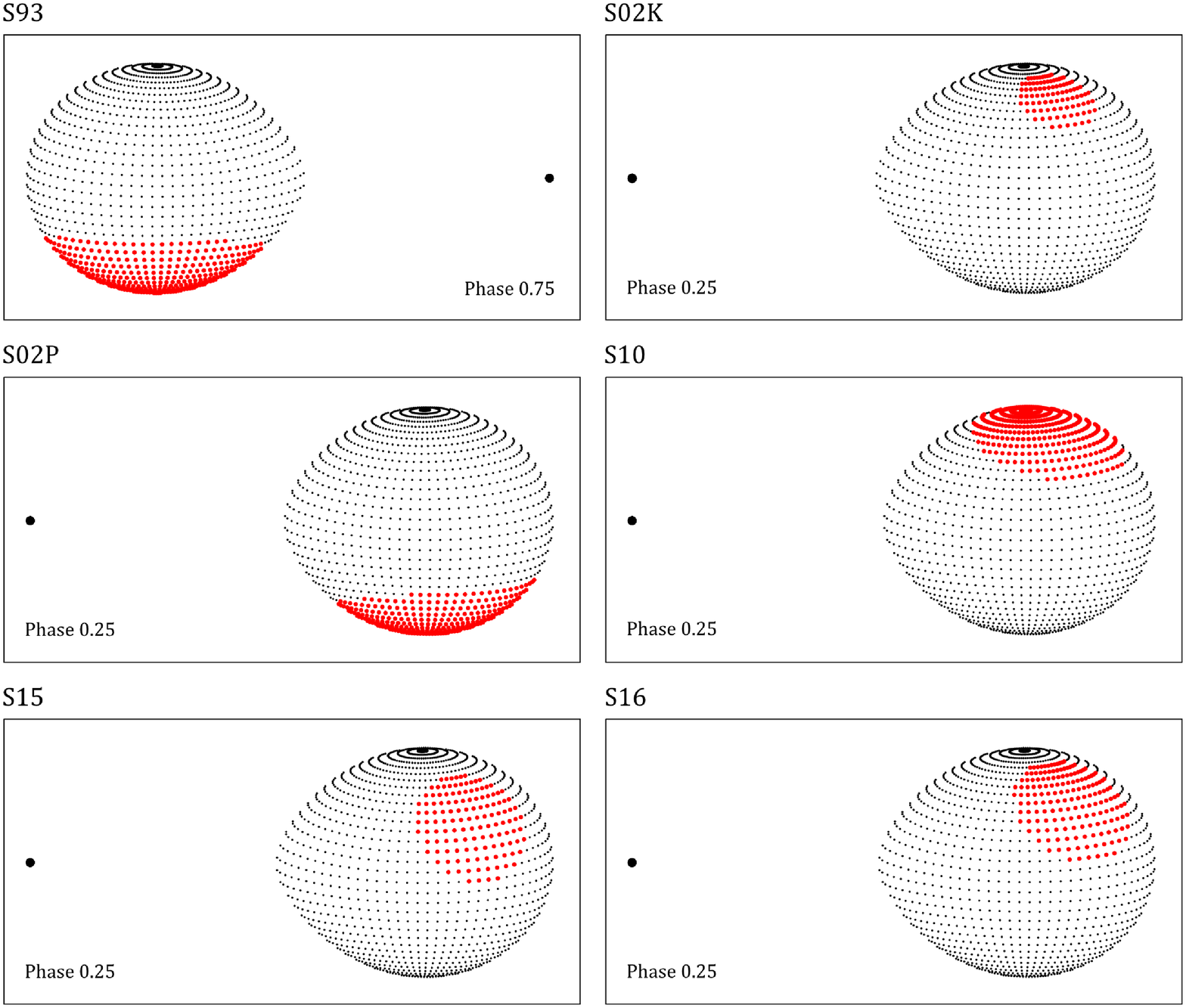}
\end{figure}

The behavior of active longitudes is summarized in Table \ref{tRes}, with seasonal and final values of other optimized parameters. Fig. \ref{fLon} shows the dependence of $\chi^2$ from spot longitude, with each point representing the best-fitting model. (The $\chi^2$ values in this plot are normalized to the \textit{maximal} value achieved for the given light curve for clarity.) Fig. \ref{fContours} shows the stellar surface maps of the MD with color-coded $\chi^2$ values at each node of spot coordinate grid. Darker colors correspond to better-fitting models.


\begin{table}
\caption{Seasonal solutions.}
\vspace{10pt}
\label{tRes}
\centering
\begin{footnotesize}
\begin{tabular}{lrrrrrrrrr}
\tableline\tableline
Season		& LCs & $\chi^2_n$ & $\lambda\ [^{\circ}]$ & $\varphi\ [^{\circ}]$ & $\theta\ [^{\circ}]$ & $i\ [^{\circ}]$ & $T_{\rm MD}\ {\rm [K]}$ & $F_{\rm MD}$ & $F_{\rm WD}$					\\
\tableline
\noalign{\smallskip}
S93		& 3	& 3.145	& 280	& -80	& 62$\pm$  8	& 78.5$\pm$1.3	& 3260$\pm$ 220	& 0.981$\pm$0.002	& 0.020$\pm$0.001	\\
S02K	& 2	& 2.118	&  60	&  55	& 21$\pm$  1	& 77.6$\pm$0.1	& 3100$\pm$  20	& 0.982$\pm$0.003	& 0.021$\pm$0.001	\\
S02P	& 3	& 7.011	&  45	& -80	& 57$\pm$ 11	& 78.0$\pm$1.1	& 3160$\pm$  80	& 0.963$\pm$0.014	& 0.021$\pm$0.001	\\
S10		& 3	& 4.561	&  50	&  70	& 41$\pm$  7	& 77.9$\pm$0.3	& 3290$\pm$ 110	& 0.969$\pm$0.011	& 0.021$\pm$0.001	\\
S15		& 4	& 4.294	&  60	&  30	& 29$\pm$  3	& 77.6$\pm$1.3	& 3190$\pm$ 170	& 0.979$\pm$0.012	& 0.021$\pm$0.002	\\
S16		& 4	& 5.110	&  55	&  40	& 31$\pm$  3	& 78.0$\pm$1.4	& 3180$\pm$ 180	& 0.978$\pm$0.014	& 0.021$\pm$0.002	\\
\hline
		&	&		&		&		&
		& \textbf{ 77.9$\pm$1.4}	
		& \textbf{ 3200$\pm$220} 
		& \textbf{0.975$\pm$0.014}	
		& \textbf{0.021$\pm$0.002}	\\
\end{tabular}
\tablecomments{Seasonal solutions obtained in the procedure described in Section \ref{sMod}. We report the values of optimized parameters averaged over filter-specific solutions; the uncertainties correspond to the value ranges as $\frac{1}{2}(max - min)$. The last row contains final averages suitable for comparison with other studies of QS Vir (e.g. with values quoted in Table \ref{tAdjust}). Final uncertainties are estimated by maximizing seasonal uncertainties.}
\end{footnotesize}
\end{table}

Figure \ref{fLon} demonstrates that spot longitudes are uniquely determined by light curves. In Fig. \ref{fContours}, feasible spot locations coincide with the dark areas corresponding to well-fitting models; it is not surprising that these areas are most sharply defined in red filters, where the contribution of MD's radiation is significant or dominant. The areas are shaped as tall rectangles, with width and height being representative of the uncertainties in determination of longitude and latitude, respectively. The white lines indicate the longitudes of seasonal solutions.


Spot latitudes are poorly constrained by the light curves regardless of filter. Other than small, likely random variations, the goodness of fit is effectively constant over the entire range of latitudes. This is hardly surprising. Weak selection of spot latitudes in light curve modeling of eclipsing binaries is a geometrical inevitability, especially when the eclipse can only hide a tiny portion of the spotted component's surface at a time, as is the case with QS Vir. Thus, the reported latitudes are provisional.

The final location and size of the spot on the MD in different seasons can be seen in Fig. \ref{fSpots}. The spot appears at high latitudes in seasons S93 through S10, and at moderate latitudes in seasons S15 and S16. Spots on magnetically active members of close binaries are expected to develop at high latitudes due to rapid rotation caused by spin-orbital synchronization \citep{solanki92}. In seasons S93 and S02P the spot is extensive and covers the south pole. However, in season S02K (which is nearly simultaneous with season S02P), we find a small spot near the north pole. This inconsistency is another symptom of low spatial resolution provided by the eclipse by the WD: a large polar spot visible in all orbital phases, but with more coverage in the first quadrature will produce a similar (possibly indistinguishable) asymmetry in the light curve as a small spot at moderate to high latitudes that's only visible in the first quadrature. This doesn't affect the determination of spot longitude, which is well-constrained by the shape of the light curve (see discussion in Section \ref{ssMigration}).


\subsection{Comparison with other studies of spots in QS Vir}
\label{ssComp}

Due to the prominent asymmetry in the light curves of QS Vir in red filters, the presence of spots on the active MD component was discussed in almost every publication dealing with this object. No model can adequately reproduce its light curves without the inclusion of one or more light or dark spots. But an analysis of spot behavior through detailed light curve modeling has never been attempted prior to this work.

\citet{ribeiro10} applied an image reconstruction method to the light curves observed by \citet{donoghue03} and \citet{kawka02} (data referred to as seasons S93 and S02K in this work) and obtained surface brightness maps of the MD that feature a total of four active regions: two dark spots and two bright spots. Apart from a small phase shift, these features look essentially the same in seasons S93 and S02K even though the shape of the light curves in the $R$ filter is markedly different, with the left-hand side maximum being the higher in S93, and the lower in S02K (Fig. 2 in \citealt{ribeiro10}; compare Figs. \ref{fS93} and \ref{fS02K} in present paper and see discussion in Section \ref{ssMigration}). A direct comparison between these surface maps and our spotted models would be difficult because we only work with one dark spot; moreover, our experiments with modeling additional spots convince us that the light curves of QS Vir do not contain enough information to extract the locations or even establish the existence of as many as four spots reliably. As we showed in Section \ref{ssCount}, even two spots are more than is required to adequately model the light curves.

\citetalias{parsons16} performed a surface brightness reconstruction using high-resolution time-resolved spectroscopy. Roche tomograms (Fig. 13 in \citetalias{parsons16}) indicate the presence of three prominent, irregular dark areas on the MD, labeled 'A', 'B' and 'C', that look about equally dark. Feature C is located at approximately the same place as the single spot in our models for seasons S02K, S10, S15 and S16.

As for features A and B, it is possible that they are accounted for by the gravity darkening in our model. Feature B is centered on the 'nose' of the MD (near the Lagrange point $L_1$), and that is the coolest region of a tidally deformed star. Feature A is located at the 'rear' of the MD (near $L_2$), the next coolest region. The temperature difference between the polar and frontal regions of the MD in our model is already about 200 K, of the same order of magnitude as the temperature difference between the spotted and unspotted stellar surface (taken to be about 300 K, as discussed in Section \ref{sMod}). Modeling spots in these locations on top of the existing temperature variation would be superfluous, especially since such spots don't contribute significantly to the asymmetries in the light curve and can't be reliably parametrized through light curve modeling. Therefore, our single-spot model does not necessarily contradict the findings of the tomographic study by \citetalias{parsons16}.

\subsection{Spot migration between 1993 and 2002}
\label{ssMigration}

Our analysis shows that the majority of spot activity (at least that part which is responsible for light curve asymmetry) has been located at and around the active longitude of $60^{\circ}$ during the past fifteen years (seasons S02K-S02P to S16), but was located at the active longitude of $280^{\circ}$ ten years prior to the earliest of those observations (during season S93).

\begin{figure}
\caption{The variation of the light curve asymmetry with time. Max I corresponds to the magnitude at maximum brightness near phase 0.25, and Max II near phase 0.75. The value shown is the average across the filters of the indicated season.}
\label{fOConnell}
\medskip
\centering
\includegraphics[width=\textwidth]{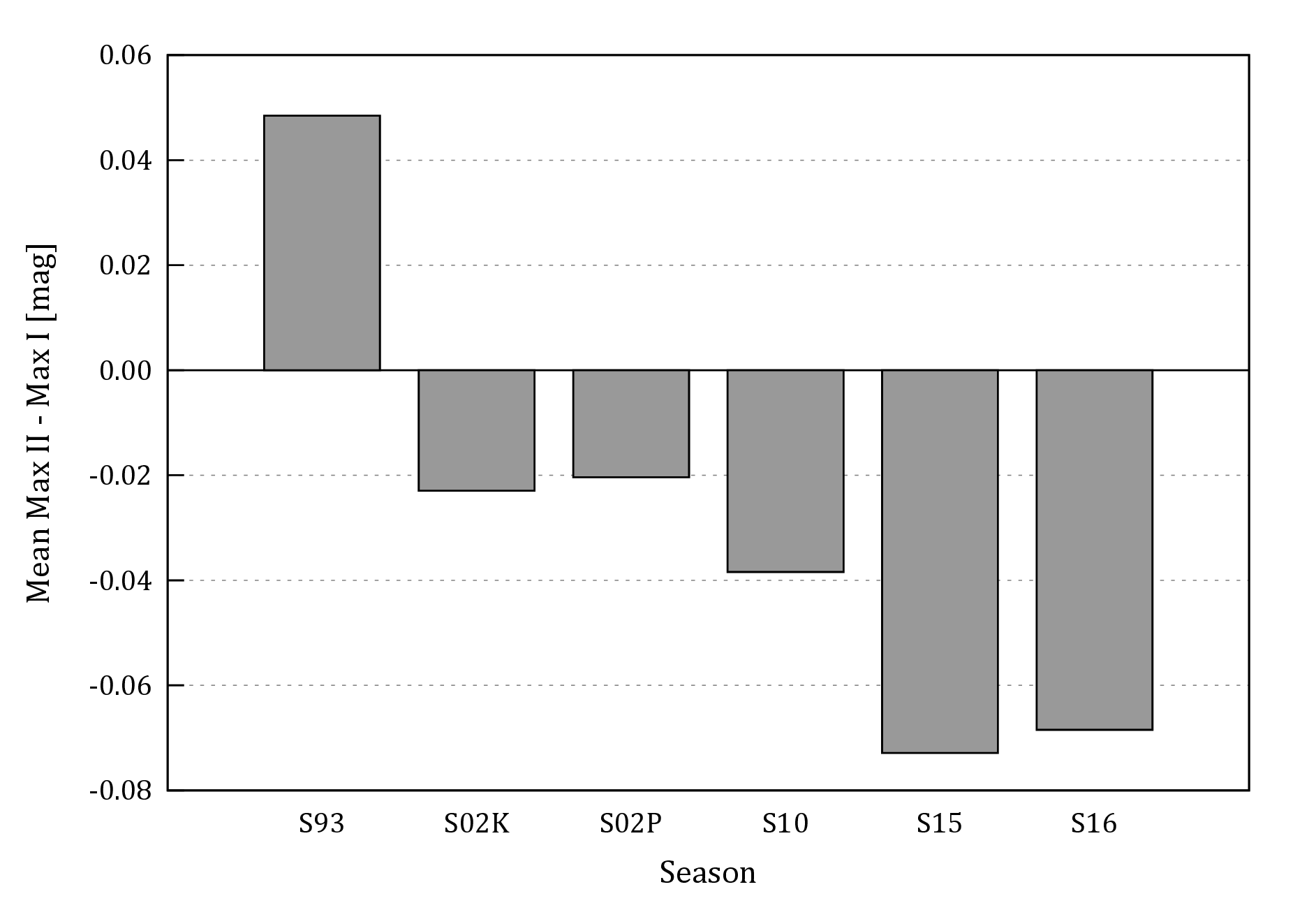}
\end{figure}

This is evident from the available data even without an elaboration on the number and precise parameters of the spots. In S93, the asymmetry of the light curve is such that the maximum near phase 0.25 is brighter than the one at phase 0.75; but in all other seasons it is just the opposite: the maximum at phase 0.75 is the brighter one. To quantify this observation, we denote the light curve maximum that follows the eclipse of the WD as $MaxI$ (near phase 0.25), and the one preceding it as $MaxII$ (near phase 0.75). In Fig. \ref{fOConnell}, we plot their difference in magnitudes, $MaxII - MaxI$, averaged across all available filters, for each season in chronological order. Clearly, that side od QS Vir which we see at phase 0.25 was brighter than the opposite side in S93 -- but darker in all other seasons. In the context of the spotted model, this means that spot coverage was more pronounced between longitudes $180^{\circ}$ and $360^{\circ}$ in S93, and between longitudes $0^{\circ}$ and $180^{\circ}$ after it. This is reflected in the results of our modeling.

While the difference between the active longitudes pinpointed by our analysis is not exactly $180^{\circ}$, we believe a flip-flop event \citep{berdyugina05, korhonen01} might be a reasonable explanation for the migration. Given the indications that both components of QS Vir might be magnetic \citep{matranga12, parsons16}, it is likely that the magnetic field of the MD has a complex structure and activity cycles. Regular long-term observations of QS Vir should be undertaken to investigate if the change in active longitudes between seasons S93 and S02 is part of a flip-flop cycle, which might further be connected to seemingly cyclic variations of the orbital period \citep{bours16}. 

\subsection{Orbital and physical parameters of QS Vir}

The other parameters of QS Vir estimated by our light curve models are in good agreement with values derived in previous studies. Our solutions give a slightly higher effective temperature of the MD on average than that estimated by \citetalias{parsons16}, but the difference ($\Delta \rm T \approx$ 100 K) is within the uncertainties usually associated with determining the effective temperature from color index. 

Final average values of orbital inclination, effective temperature of the MD, and the filling factors of both components can be found in the bottom row of Table \ref{tRes}. The absolute parameters of QS Vir were calculated for every season separately using the spectroscopic elements from \citetalias{parsons16}; we list their final average values in Table \ref{tAbs}.


\begin{table}
\caption{Absolute parameters of QS Vir.}
\vspace{10pt}
\label{tAbs}
\begin{center}
\begin{tabular}{lllllc} 
\tableline\tableline
Parameter								& Value				\\	
\tableline
\noalign{\smallskip}
${\cal L}_{\rm WD} {\rm [L_{\odot}]}$	& 0.0043	$\pm$ 0.0003	\\
${\cal L}_{\rm MD} {\rm [L_{\odot}]}$	& 0.014		$\pm$ 0.002	\\
${\cal M}_{\rm WD} {\rm [M_{\odot}]}$	& 0.783		$\pm$ 0.001	\\
${\cal M}_{\rm MD} {\rm [M_{\odot}]}$	& 0.383		$\pm$ 0.001	\\
${\cal R}_{\rm WD} {\rm [R_{\odot}]}$	& 0.011		$\pm$ 0.001 \\
${\cal R}_{\rm MD} {\rm [R_{\odot}]}$	& 0.387	    $\pm$ 0.005 \\
${\rm log}(g)_{\rm WD}$					& 8.26 		$\pm$ 0.03	\\
${\rm log}(g)_{\rm MD}$					& 4.85 		$\pm$ 0.02	\\
$\Omega_{\rm WD}$						& 116		$\pm$ 3	\\
$\Omega_{\rm MD}$						& 2.90		$\pm$ 0.02	\\
\tableline
\end{tabular}
\tablecomments{${\cal  L}$, ${\cal M}$ and ${\cal R}$ are the luminosities, stellar masses and radii in Solar units; ${\rm log}(g)$ is the logarithm (base 10) of effective gravitational acceleration in cgs, and $\Omega$ is the dimensionless surface potential. Uncertainties correspond to season-to-season variations as $\frac{1}{2}(max - min)$.}
\end{center}
\end{table}

\section{Conclusions}
\label{sCon}

We observed the post-common envelope white dwarf + M dwarf close binary QS Vir in 2015 and 2016 in \textit{BVRI} filters and obtained light curves that could not be modeled to our satisfaction following the usual approach of simultaneous fitting in all filters. When we applied our model to archival light curves observed in 1993, 2002 and 2010, we found that they could not be fitted simultaneously either. To cope with the issue, we devised a grid-search procedure that can effectively map spots on the surface of the MD. This led us to the following conclusions:

\begin{itemize}
\item{A single dark spot on the MD is sufficient to explain the asymmetries in the light curves across all filters and all observational seasons. Additional spots do not significantly improve the fit of the model to the data. While this does not disprove the existence of multiple spotted areas, it does indicate that light curves alone do not contain enough information to reliably parametrize more than one spot.}
\item{The longitude of a single spot can be determined reliably and uniquely as a result of our analysis. The latitude can not.}
\item{The single dark spot in our final models has been a stable feature at $60^{\circ}$ longitude for the past 15 years, after a major migration from $280^{\circ}$ longitude that happened between 1993 and 2002. The difference between these active longitudes is near enough to $180^{\circ}$ to suggest a flip-flop event as its likely cause.}
\end{itemize}

Further research into the behavior of spots on QS Vir would benefit immensely from regular photometric and spectroscopic observations. A long-term study could reveal if the flip-flop event between 1993 and 2002 was a part of a magnetic cycle, and whether such a cycle could be related to the unexplained period variations.

\acknowledgements

We extend our gratitude to S.G. Parsons and M.C.P. Bours for kindly sharing with us the observational data and research results that greatly increased the quality of this work. We also thank the anonymous referee for the careful reading of the manuscript and constructive suggestions. During the research presented in this paper, the authors received support from the Ministry of Education, Science and Technological Development of Republic of Serbia (project No. 176004), and the Science and Technological Development Fund, Ministry for Scientific Research of Egypt (project ID 1335). We gratefully acknowledge the use of the Simbad database ({\it http://simbad.u-strasbg.fr/simbad/}), operated at the CDS, Strasbourg, France, and NASA's Astrophysics Data System Bibliographic Services ({\it http://adsabs.harvard.edu/}).

\end{document}